\documentclass[conference]{IEEEtran}

\usepackage{amsmath,amssymb,amsfonts}
\usepackage{algorithmic}
\usepackage{graphicx}
\usepackage{textcomp}
\usepackage{url}

\usepackage{xargs}
\usepackage{bm}
\usepackage{tabularx}
\usepackage{booktabs}
\usepackage{longtable}
\usepackage{multirow}
\usepackage{makecell}
\usepackage{array}
\usepackage[super]{nth}
\usepackage{float}
\usepackage{wrapfig}
\usepackage{multicol}

\begin{document}
\title{Unstable Throughput: When the Difficulty Algorithm Breaks}

%\titlerunning{Unstable Throughput}
% If the paper title is too long for the running head, you can set
% an abbreviated paper title here

\author{\IEEEauthorblockN{Dragos I. Ilie, Sam M. Werner, Iain D. Stewart and
William J. Knottenbelt}
\IEEEauthorblockA{Imperial College London\\
London, United Kingdom\\
}}
\IEEEoverridecommandlockouts
\IEEEpubid{\makebox[\columnwidth]{978-0-7381-1420-0/21/\$31.00~\copyright2021 IEEE\hfill} \hspace{\columnsep}\makebox[\columnwidth]{ }}
\maketitle

%%% REMOVE BEFORE SUBMISSION !!! 
% \thispagestyle{plain}
% \pagestyle{plain}

\begin{abstract}
In Proof-of-Work blockchains, difficulty algorithms serve the crucial purpose of maintaining a stable transaction throughput by dynamically adjusting the block difficulty in response to the miners' constantly changing computational power. 
Blockchains that may experience severe hash rate fluctuations need difficulty algorithms that quickly adapt the mining difficulty.
However, without careful design, the system could be gamed by miners using coin-hopping strategies to manipulate the block difficulty for profit.
Such miner behavior results in an unreliable system due to the unstable processing of transactions.

We provide an empirical analysis of how Bitcoin Cash's difficulty algorithm design leads to cyclicality in block solve times as a consequence of a positive feedback loop.
In response, we mathematically derive a difficulty algorithm using a negative exponential filter which prohibits the formation of positive feedback and exhibits additional desirable properties, such as history agnosticism.
We compare the described algorithm to that of Bitcoin Cash in a simulated mining environment and verify that the former would eliminate the severe oscillations in transaction throughput.

\end{abstract}

\begin{IEEEkeywords}
Blockchain, Bitcoin, Bitcoin Cash, Difficulty Algorithm, Mining, Coin-Hopping
\end{IEEEkeywords}

\section{Introduction}
\label{ref:introduction}

Proof-of-Work (PoW) blockchains offer a decentralized mechanism for recording data in a trustless and immutable manner. 
% Participants encode the exchange of value or information in transactions and broadcast them through a peer-to-peer network. 
% Miners, the backbone of the system, aggregate transactions in data structures called blocks and append them to the blockchain.
To reach consensus over the ordering and validity of blocks, miners participate in a leader election process by solving a computationally intensive puzzle named Proof-of-Work. 
The first miner to find a valid solution appends a block and receives a reward for the invested computational effort.
To ensure stable transaction throughput, a difficulty algorithm (DA) adjusts the difficulty of the PoW puzzle in response to changes in the miners' computational power.
However, without careful design the DA can expose vulnerabilities, which when exploited by miners, lead to inappropriate difficulty levels and thus patterns of instability in the transaction throughput.
In general, this issue arises in blockchains that lack a consistent amount of computational power due to some miners directing their resources towards other blockchains especially as profitability varies.
For instance, such patterns have been observed in Bitcoin Cash (BCH)~\cite{BitmexResearch,bitcoinabc}, the cryptocurrency with the \nth{7} highest market capitalization of \$$5$bn\footnote{Data obtained from: \url{https://coinmarketcap.com}. Accessed: 22-11-2020.}.
Its developers have announced that fixing the high variations in block solve times is one of their main development goals for 2020~\cite{web:bitcoinabc}.
Two proposals~\cite{aserti3proposal,sechet2020announcing} have been put forward to replace BCH's DA known as {\tt cw-144}.
In this paper, we present a negative exponential filter DA (NEFDA) which is similar to these proposals and is explicitly referenced\footnote{The reference has been made to an earlier pre-printed version of this paper.} by one of them~\cite{aserti3proposal}.
Additionally, we define desirable properties exhibited by NEFDA and prove their benefits by comparing the performance of NEFDA with {\tt cw-144} in a simulated mining scenario.
% On 15 November 2020, the two competing proposals were deployed via hard forks causing a chain split~\cite{forknews}.
% As our work was carried out prior to this fork, the remainder of this paper refers to {\tt cw-144} as ``BCH's current DA", which is technically no longer the case.

% \paragraph*{Contributions}
% In this paper, we derive from first principles a DA designed to stabilize transaction throughput even in blockchains without consistent computational power.
% To this end, it adjusts the difficulty after every block using exponential smoothing, a popular approach for removing noise in time series data.
% We provide an empirical analysis on {\tt cw-144} and investigate its inherent vulnerabilities.
% We discover that economically rational (i.e.\ profit-seeking) miner behavior leads to severe instabilities in transaction throughput due to a positive feedback loop in block solve times resulting from the design of the DA.
% We present desirable properties of the derived DA and show how they remove the positive feedback loop and demonstrate through simulations how it performs under different scenarios and compare it to {\tt cw-144}.
% We find that the derived DA would be an improvement and suggest that it could be applicable to any PoW blockchain when configured appropriately.

\section{Related Work}
\label{ref:related_work}
The most extensive body of difficulty algorithm research has been done by the pseudonym zawy12, who provides a comprehensive overview of various difficulty algorithms in~\cite{zawy12SummaryDA}.
He examines the difficulty instabilities in BCH in~\cite{zawy122019bchNewDA} and simulates the performance of various DAs, including ASERT~\cite{lundeberg2020staticV2}\footnote{We have become aware of ASERT, which is essentially equivalent to NEFDA, after receiving the unpublished work~\cite{lundeberg2019static} of Mark B. Lundeberg from zawy12.} and EMA~\cite{zawy12UsingEMA,eliosoff2017ema,eliosoff2018wema}, which is an approximation of ASERT that avoids the computation of exponentials.
Our work differentiates by providing a formal derivation of NEFDA starting from first principles and an outline of the desirable properties achieved.

\section{Background}
\label{ref:background}
\label{ref:prelim_daa}
In this section, we introduce readers to the difficulty algorithms used in Bitcoin (BTC)~\cite{nakamoto2008bitcoin} and BCH.

A \textit{difficulty algorithm} (DA) is a fundamental component of PoW blockchains as it ensures a stable transaction throughput by adjusting the hardness of generating a PoW solution.
% A DA is responsible for ensuring stable block times in periods of hash rate oscillations caused by miners joining and leaving the network.
% Failing to ensure an appropriate difficulty could result in either short time periods with many blocks being found, or long time periods with very few blocks, which results in a highly variable response time for blockchain transactors.
An omniscient DA, with knowledge of the real world hash rate, would be able to compute the difficulty of the next block by simply multiplying the current hash rate with the ideal inter-block time (e.g. $10$ minutes). 
However, computing the difficulty of a block must be deterministic and based on data from previous blocks s.t.\ individual nodes can perform the computation independently and agree on the same results.
Therefore, DAs estimate the current hash rate based on the difficulties and solve times of blocks in the recent past.
The extent to which a DA is able to minimize the lag between the actual hash rate and the estimated one is regarded as the reactiveness of the algorithm.
Blockchains with relatively stable hash rate can afford to use a less reactive DA to reduce volatility in difficulty, allowing miners to predict expected rewards over near-term time scales.
The frequency of difficulty adjustments also affects the reactiveness of a DA.

\subsection{BTC Difficulty Algorithm}
In BTC, the new difficulty $D^\prime$ is updated every $2\,016$ blocks (approx. $2$ weeks) based on the previous difficulty $D$, using the following formula:
\begin{align}
    D^\prime &= D \cdot \text{max}\left(\text{min}\left(\frac{2\,016 \cdot T}{T_A}, 4\right), \frac{1}{4}\right)
\end{align}
where $T$ is the ideal inter-block time (i.e.\ $10$ minutes) and $T_A$ is the time it actually took to mine the last $2\,016$ blocks.

\subsection{BCH Difficulty Algorithm}
BCH's DA, referred to as \texttt{cw-144}, attempted to increase responsiveness to both effluxes and influxes of hash rate by performing difficulty adjustments on a per-block basis.
The difficulty $D$ of a new block is derived from the estimated hash rate, $\widehat{H}$, and the ideal inter-block time, $T$.
To this end, $\widehat{H}$ is computed using a simple moving average with a sample size of approx.\ $144$ blocks from $B_{\mathit{start}}$ to $B_{\mathit{end}}$.
% To mitigate situations when the block timestamps are out-of-order, the bounds of the sliding window over which the average is computed are derived using the median timestamp of $3$ blocks.
% Thus, the block at which the window starts, $B_{\mathit{start}}$, is the block with the median timestamp out of blocks $144, 145$, and $146$ in the past.
% Similarly, the window ends at block, $B_{end}$, with the median timestamp of the $3$ most recent blocks.
%From these two blocks 
The DA computes $CW$, the amount of chain work that was performed between these two blocks, as the sum of difficulties of all blocks in the interval $[B_{\mathit{start}}, B_{\mathit{end}}]$.
The estimated hash rate is: $\widehat{H} = CW / T_A$, where $T_A$ is the actual time elapsed between $B_{\mathit{start}}$ and $B_{\mathit{end}}$, capped in the interval from half a day to $2$ days to prevent difficulty changing too abruptly.
Thus, the equation for the new difficulty is:
\begin{equation}
    D = \widehat{H} \cdot T =  \frac{\sum\limits_{i=\mathit{start}}^{\mathit{end}} \text{diff}(B_i)}{T_A} \cdot T
\label{eq:BCH_DA}
\end{equation}

\section{Empirical Analysis of BCH's DA}
% \label{ref:empirical}
% % \todo[inline]{Data examined:
% % For the empirical analysis data has been examined for these block number periods:
% % BTC: 478558 (Aug. 1, 2017) -- 621220 (Mar. 11, 2020: 3pm)
% % BCH: 478558 (Aug. 1, 2017) -- 625989 (Mar. 11, 2020: 3pm)
% % }
In this section, we provide an empirical analysis of issues stemming from the use of {\tt cw-144} in BCH.
% Note that when we refer to BCH, we are referring to the Bitcoin ABC~\cite{web:bitcoinabc} full node implementation. 

% \subsection{BCH's Difficulty Algorithm}
% When the BTC--BCH fork occurred on \nth{1} August 2017, BCH kept BTC's PoW and DA, but also added an \textit{Emergency Difficulty Algorithm} (EDA) to protect against coin-hopping strategies. 
% The EDA would cause the difficulty to drop by $20$\% if the difference between $6$ successive block timestamps exceeds $12$ hours~\cite{aggarwal2019structural}.
% However, it soon became apparent that this difficulty adjustment mechanism did not fulfill its objective.
% Miners would stop mining BCH in order to cause consecutive $20$\% drops in the difficulty, which only adjusted back upwards every $2\,016$ blocks.
% Once the difficulty was sufficiently low, miners would switch back to mining BCH and produce many blocks at very low difficulty until the end of the $2\,016$ blocks window.

% The combination of BTC's DA and EDA was replaced in BCH on \nth{13} November 2017 with a new DA, referred to as \texttt{cw-144}.

\subsection{Oscillations in Number of Blocks Mined per Hour}
\begin{figure}[H]
\centering
\includegraphics[]{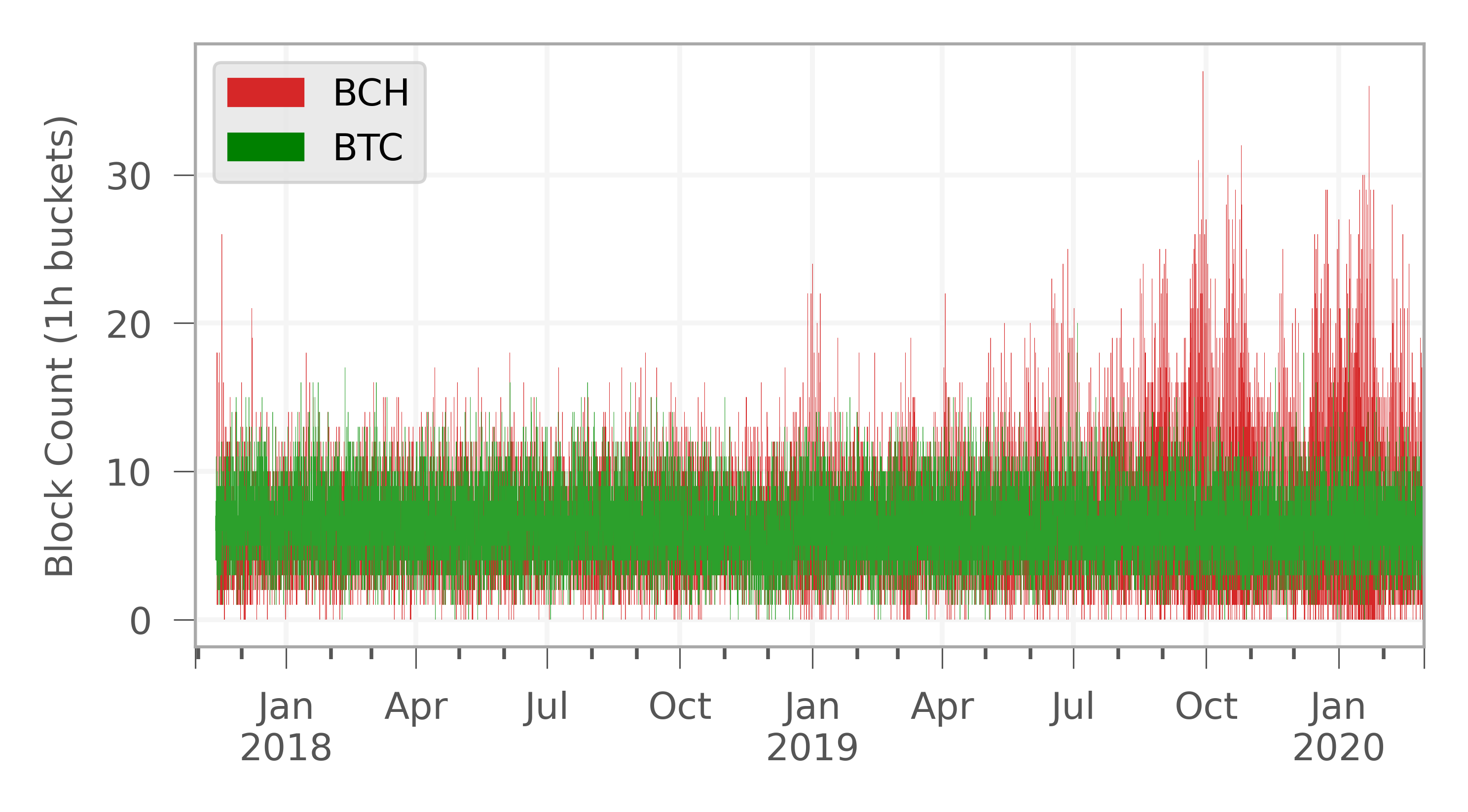}
\caption{Number of blocks mined per hour in BTC and BCH.}
\label{fig:total_blocks_bch_btc_1H}
\end{figure}

As intended, \texttt{cw-144} achieves a daily average solve time of $10$ minutes.
This gives the superficial impression of performing well in terms of stable throughput; however, certain patterns in the distribution of blocks within a day emerge.
From Figure~\ref{fig:total_blocks_bch_btc_1H} it can be seen that the oscillations in number of blocks mined per hour are notably more severe in BCH than in BTC.
Especially during the later months, it is evident that BCH exhibited more $1$ hour periods with either many blocks mined or none.

% As the number of blocks mined in an hour, $K$, should ideally follow a Poisson distribution with rate parameter $\lambda=6$ blocks, we can compute the expected probability of mining exactly $k$ blocks in one hour as:
% \begin{equation}
%     P(K = k) =\frac{6^{k}}{e^{6}{k!}} \label{eq:blocksperhour}
% \end{equation}

% We compare these ideal values with empirical results from BCH and BTC in Figure~\ref{fig:prob_mass_fun}.
% For reference, it can be seen that in BTC the probabilities closely resemble those of a Poisson process.
% In contrast, BCH shows significant deviations from the ideal distribution during the period in which the EDA was active.
% After abandoning the EDA, BCH has indeed shifted towards the Poisson distribution, but a skew on the left and right tails remains.
% This is in line with the aforementioned observations of a more unstable transaction throughput in BCH, as shown in Figure~\ref{fig:total_blocks_bch_btc_1H}.
% \begin{figure}[H]
%     \centering
%     \includegraphics[width=.47\textwidth]{figures/prob_mass_fun.pdf}
%     \caption{The probabilities of mining exactly $k$ blocks in a one-hour period in BTC and BCH (pre and post EDA).}
%     \label{fig:prob_mass_fun}
% \end{figure}

\subsection{Positive Feedback Loop in Simple Moving Averages}
\label{ref:sma_instability}
The observed instability in transaction throughput can be explained by a positive feedback loop that stems from a combination of two factors: the use of a simple moving average and miners' economically rational behavior.

\begin{figure}[H]
\centering
\includegraphics[width=0.47\textwidth]{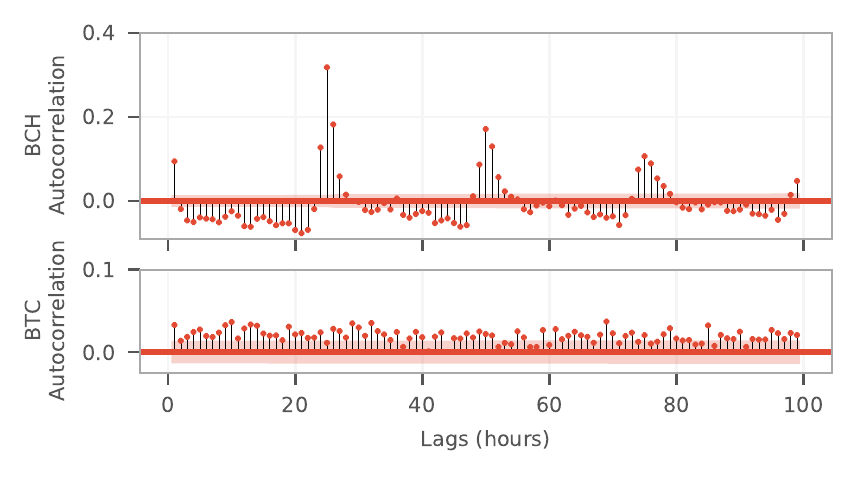}
\caption{The autocorrelation in number of blocks mined per hour in BCH and BTC since \texttt{cw-144} was deployed.
% BTC block numbers: $494120$ -- $621220$; BCH block numbers: $504067$ -- $625989$.
}
\label{fig:acf_bch_btc_1h_blocks}
\end{figure}

From equation~\eqref{eq:BCH_DA} it is apparent that \texttt{cw-144} relies (in part) on the relationship of inverse proportionality between $T_A$ and the estimated hash rate, $\widehat{H}$.
The same relationship exists between the hash rate fluctuations and the solve times of newly mined blocks.
As new solve times are added to $T_A$, the result of these two relations is that $\widehat{H}$ is adjusted directly proportional to the actual hash rate change.
However, the oversight of this DA is that using a simple moving average implies solve times falling off the window (subtracted from $T_A$) have an equal weight in the computation of $\widehat{H}$.
Short solve times $144$ blocks in the past cause a relative increase in $T_A$ which yields a lower-than-expected $\widehat{H}$. 
Similarly, long solve times falling off the window imply a relative decrease in $T_A$ and therefore produce a higher $\widehat{H}$.
This influence constitutes positive feedback that results in correlation between solve times $144$ blocks ($24$ hours) apart, as can be seen in Figure~\ref{fig:acf_bch_btc_1h_blocks}.

The second factor that contributes towards the positive feedback loop is the miners' behavior as they try to maximize profit by engaging in coin-hopping~\cite{meshkov2017short,kiraly2018profitability,kwon2019bitcoin}.
Assume BCH experiences an increase in profitability which incentivizes a group of coin-hopping miners $M_{\mathit{CH}}$ to switch their computational power towards BCH. 
This causes an increase in hash rate and consequently a series of blocks with short solve times.
As the difficulty adjusts upwards, BCH's profitability drops until $M_{\mathit{CH}}$ leave the network and the hash rate returns to its original value.
However, the difficulty is now too high for the network, so a series of blocks with long solve times is produced.
The positive feedback of blocks falling off the window, causes this pattern of short solve times followed by long solve times to repeat forming a positive feedback loop.
As can be seen from Figure~\ref{fig:acf_bch_btc_1h_blocks}, only BCH experiences such a feedback loop because its base hash rate is approx.\ $3\%$ of BTC's hash rate so coin-hopping miners have a much more significant impact.
This phenomenon has also been examined by~\cite{zawy122018sma,zawy122019bchNewDA,reddit:BCH,jtoomim2020upgrade}.

\section{Negative Exponential Filter Difficulty Algorithm}
\label{ref:model}

In this section, we mathematically derive a negative exponential filter difficulty algorithm (NEFDA) based on a common technique for removing noise from time series data known as exponential smoothing. 
Throughout we make use of the following notation:
\begin{align*}
    D_i & \leftarrow \text{difficulty of block } i & \\
    t_i & \leftarrow \text{time of block } i & \\
    st_i & \leftarrow \text{solve time of block } i: st_i = t_i - t_{i-1} & \\
    T & \leftarrow \text{ideal block solve time (e.g. 10 minutes)} & \\
    S & \leftarrow \text{decay/smoothing factor (see Section~\ref{sec:smoothingfactor})}\\
    \widehat{H}_i & \leftarrow \text{estimated hash rate at block } i
\end{align*}
For simplicity, index $0$ refers to the block at which the new DA is deployed, while index $n$ refers to the next block to be appended. 
Thus, $t_n$ and $\widehat{H}_n$ represent the current time and network hash rate, respectively.
NEFDA uses real time targeting (RTT), i.e.\ the difficulty $D_n$ of the block that is being mined dynamically adjusts as time passes.
$$D_n = D_{0} e^{\tfrac{t_0 + n T - t_n}{S}}$$
Considering RTT is not a popular technique, we argue at length for its safety in Section~\ref{sec:honest_timestamps}.
In the remainder of this section, we show how this formula is derived from first principles.

% Other works~\cite{lundeberg2020staticV2,zawy12SummaryDA} refer to similar formulations of this algorithm as ``absolutely scheduled exponentially rising targets'' (ASERT).

\subsubsection{Estimating Current Hash Rate}
Difficulty algorithms are in the business of estimating the current network hash rate, $\widehat{H}_n$. 
As the actual function of hash rate cannot be known at any given time we rely on sampling when information is available, i.e.\ when blocks are mined.
On average, the difficulty $D_i$ represents the number of hashes computed throughout the interval $(t_{i-1}, t_i]$.
Approximating that $D_i$ hashes are computed at time $t_i$ we can estimate the current hash rate $\widehat{H}_n$ using exponential smoothing over the series of block difficulties, i.e.\ by taking their exponentially weighted average.
\begin{align}
    \widehat{H}_n = \frac{\sum\limits_{i = 0}^{n-1} D_i   e^{\tfrac{t_i - t_n}{S}}}{\int\limits_{-\infty}^{0} e^{\tfrac{x}{S}} dx} = \dfrac{1}{S}\sum\limits_{i = 0}^{n-1} D_i   e^{\tfrac{t_i - t_n}{S}}
    \label{eq:hnintegral}
\end{align}

\subsubsection{Difficulty Computation}
Therefore, the difficulty $D_n$ of the next block is:
\begin{align}
D_n &= T \cdot \widehat{H}_n = \dfrac{T}{S}\sum\limits_{i = 0}^{n-1} D_i   e^{\tfrac{t_i - t_n}{S}} \label{eq:diffn}\\
    &= \dfrac{T}{S} \sum\limits_{i = 0}^{n-1} D_i e^{\tfrac{t_i - t_{n-1}}{S}} e^{\tfrac{t_{n-1} - t_{n}}{S}} \label{eq:extractst}\\
    &= e^{\tfrac{-st_n}{S}} \left(\dfrac{T}{S}\sum\limits_{i = 0}^{n-2} D_i e^{\tfrac{t_i - t_{n-1}}{S}} + \dfrac{T}{S} D_{n-1} \right) \label{eq:extractlastterm}\\
    &= e^{\tfrac{-st_n}{S}} \left(D_{n-1} + \dfrac{T}{S} D_{n-1} \right)\\
    &=  D_{n-1} \left(1 + \tfrac{T}{S}\right) e^{\tfrac{-st_n}{S}} \label{eq:dn-1}
\end{align}

When unwinding the recurrence relation~\eqref{eq:dn-1} all the way to $D_0$ we obtain:
\begin{align}
    D_n &= D_0 \left(1 + \tfrac{T}{S}\right)^n \prod\limits_{i = 1}^{n} e^{\tfrac{-st_i}{S}} = D_0 \left(1 + \tfrac{T}{S}\right)^n e^{\tfrac{t_0 - t_n}{S}} \label{eq:d0reduced}
\end{align}

\subsubsection{Correction}
\label{model:ecorrection}
Notice that when $T \ll S$ we can approximate $1 + T/S \approx e^{T/S}$.
In fact, this is actually a correction needed to mitigate the bias introduced when considering a discrete series of difficulties instead of the continuous function of hash rate.
To prove this, we replace the constant term: $1 + T / S$ with $c$ and compute its value when the DA operates under a simple theoretical scenario.
Specifically, we assume the hash rate remains constant for many blocks between $m$ and $n$.
Thus, we expect the average rate of change in difficulty $\overline{R} = 1$, indicating that on average the difficulty does not change.
We take the geometric mean of ratios of consecutive difficulties from block $m$ to $n$ and use equation~\eqref{eq:d0reduced} with the $c$ replacement:
\begin{align}
    \overline{R} &= \sqrt[n-m]{\prod\limits_{i = m + 1}^n \dfrac{D_i}{D_{i-1}}} = \sqrt[n-m]{\dfrac{D_n}{D_{m}}} \\
    &= \sqrt[n-m]{\dfrac{D_0 c^n e^{(t_0 - t_n)/S}}{D_0 c^m e^{(t_0 - t_m)/S}}} = c \cdot e^{\tfrac{t_m - t_n}{(n-m)S}} \label{eq:ceavg}
\end{align}
Assuming the DA is working correctly, the average solve time of blocks from $m$ to $n$ is $(t_n - t_m)/(n-m) = T$.
Replacing in equation~\eqref{eq:ceavg} we obtain: $\overline{R} = 1 = c \cdot e^{\tfrac{-T}{S}} = 1$ which implies $c = e^{T/S}$.

Therefore, the correction is indeed justified and applying it in equations~\eqref{eq:dn-1} and~\eqref{eq:d0reduced}, gives the following relative and absolute forms:
\begin{multicols}{2}
\noindent
\begin{equation}
D_n = D_{n-1} e^{\tfrac{T - st_n}{S}} \label{eq:dn-1final}
\end{equation}
\noindent
\begin{equation}
D_n = D_{0} e^{\tfrac{t_0 + n T - t_n}{S}} \label{eq:d0final}
\end{equation}
\end{multicols}

\subsection{Properties}
\subsubsection{History Agnosticism}
The distribution of blocks in a given time period does not influence the difficulty of the block currently being mined.
This property is desirable as block arrivals should be independent of each other so the difficulty of a block should not depend on the history of the chain.
Equation~\eqref{eq:d0final} shows how the difficulty at time $t_\alpha$ depends only on the blockchain height, regardless of whether blocks were mined a long time in the past, in the last hour, or equally distributed in time.

\subsubsection{Lack of Autocorrelation}
Not only does this algorithm avoid the use of a sliding window, but the lack of autocorrelation is an emergent property entailed by history agnosticism.
Sudden influxes or effluxes of hash rate may still produce temporary spikes or deserts, yet their duration will be much shorter. 
However, these will not create a positive feedback loop as the distribution of blocks in time has no influence on the future.
Therefore, the inherent negative feedback present in NEFDA is the only force acting on solve times.

\subsection{Smoothing Factor Considerations}
\label{sec:smoothingfactor}
The smoothing factor $S$ has the function of configuring the reactiveness of NEFDA by setting the maximum rate of upward adjustments for the difficulty.
More specifically, the difficulty can increase by at most a factor of $e$ in $S/T$ blocks. 
Depending on the requirements of the application, $S$ should be chosen carefully: blockchains that are expected to experience large hash rate fluctuations on a regular basis (e.g.\ BCH), should aim for smaller values of $S$ to obtain a more reactive DA, while blockchains with a relatively stable hash rate (e.g.\ BTC) can choose larger values for $S$ to reduce the difficulty's volatility.
There is no direct relationship between the smoothing factor of an exponential moving average and the sample size of the simple moving average used in \texttt{cw-144}, as their operation is considerably different, but our simulations as well as other empirical studies~\cite{zawy12SummaryDA} suggest that in order to obtain similarly stable difficulties the smoothing factor should be chosen to represent $(N+1)/2$ blocks where $N$ is the length of the sliding window used in simple moving averages.
Applying this heuristic to BCH which has a sliding window of $144$ blocks, suggests $S$ should be set at approx.\ $12$ hours.

\subsection{Real Time Targeting Considerations}
\label{sec:honest_timestamps}
Real time targeting DAs assume miners have no incentive to report incorrect timestamps.
To prove this assumption we compare NEFDA's RTT formulation with BCH's \texttt{cw-144} and argue that NEFDA reduces the incentives for timestamp manipulation.
In \texttt{cw-144} reporting a dishonest timestamp, with a value in the future, would lower the difficulty for the next $144$ blocks.
This creates short term incentives for other miners to accept the dishonest block as they also benefit from the reduced difficulty even if they are not planning to be dishonest themselves.
In contrast, NEFDA's history agnosticism implies that only the difficulty of the block with dishonest timestamp is affected, so there are no incentives for other miners to accept it.
In fact, building on a dishonest block ($B_i$) implies mining towards a difficulty that is $e^{T/S}$ times higher than that of the previous block ($B_{i-1}$).
Thus, a miner would only accept this block if it is willing to report an even higher timestamp to mitigate the increase in difficulty.
This behavior leads to an unstable chain as it could be replaced by a potentially shorter chain with more accumulated work (higher difficulty blocks), so honest miners would not risk accepting blocks with dishonest timestamps.
Only an attack supported by a majority of the hash rate would be successful, which is no different than $51$\% attacks~\cite{szalachowski2018short,boverman2011timejacking,zawy122018timestamps} that are currently possible in BCH or even BTC.

% As we are only interested in simulating miners' behavior and not the relation between specific chains, we simplify profitability computations by only comparing the DARI with its initial value and assuming a constant exchange rate as most \texttt{SHA-256} coins are highly correlated.
% Thus, we assume miners compute DARIs as the ratio between the average target of the last $6$ blocks and the initial target of the blockchain.
% We believe the following $3$ types of miners are general enough to capture real miner behavior when configured appropriately.

% \textit{Idealistic miners} allocate all their hash rate to BCH regardless of how much profitability drops; they represent the baseline hash rate: $H_B$.

% \textit{Greedy coin-hopping miners} allocate all their hash rate, $H_G$, towards BCH only when the profitability is 5\% higher than its initial value.

% \textit{Variable coin-hopping miners} allocate part of their total hash rate, $H_V$, in relation to the current profitability.
% Although this relation is not clear in reality, we consider a model based on the logistic curve.
% The intention is to emulate both the initial stage when the hash rate increases exponentially as miners realize the advantage in profitability, and the later stage when the hash rate influx gradually slows down.
% The model directs all the hash rate away from or towards BCH, if drops or increases in profitability larger than $15$\% occur.
% Otherwise, a variation $x$ between $-15$\% and $15$\% leads to a contribution of $H = H_v / (1 + e^{-6/0.15 \cdot x})$ towards the total hash rate.

\section{Simulation}
\label{ref:results} 
In this section, we empirically analyze the robustness of NEFDA, by comparing it with \texttt{cw-144} over a $100\,000$ blocks period.
We simulate the behavior of coin-hopping miners by adjusting the total hash rate in response to fluctuations in profitability.
To stress test NEFDA, we consider a rather extreme scenario where greedy and variable coin-hopping miners have hash rates $H_G = H_V = 4 \times H_B$ (the base hash rate). Greedy miners allocate all their hash rate, $H_G$, when profitability is $5\%$ higher compared to the initial value while variable miners allocate their hash rate using a logistic curve: $H = H_V / (1 + e^{-6/0.15 \cdot x})$ where $x$ represents the change in profitability.

\begin{figure}[H]
\centering
\includegraphics[trim={0 0.95cm 0 0},clip, width=0.48\textwidth]{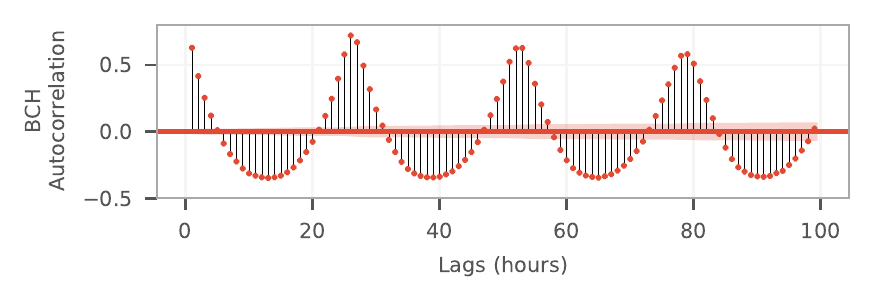}
\includegraphics[width=0.48\textwidth]{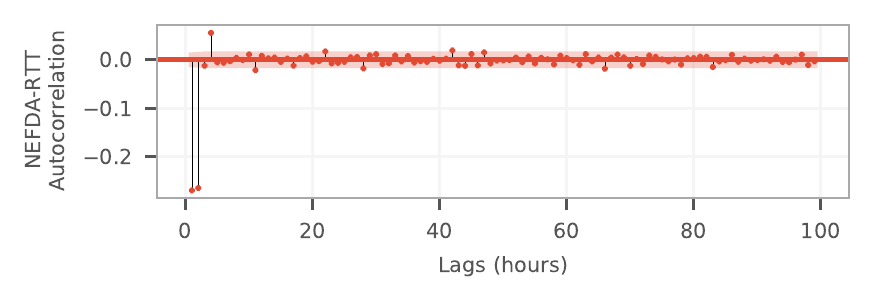}
\caption{The autocorrelation in number of blocks mined per hour in \texttt{cw-144} (top) and NEFDA (bottom)}
\label{sim:acf-rtt-bch-v4-g4-n3}
\end{figure}

A brief analysis of the average solve times: $599.97$\,s for NEFDA and $604.34$\,s for {\tt cw-144}, already reveals how NEFDA achieves a more appropriate value under this extreme scenario.
Furthermore, Figure~\ref{sim:acf-rtt-bch-v4-g4-n3} plots the autocorrelation in number of blocks mined per hour in \texttt{cw-144} and NEFDA.
For the former, a significant amount of positive autocorrelation appears at multiples of $24$ (the number of hours in BCH's sliding window).
The negative correlation between periods that are $12$ hours apart is expected considering the effect of averaging over a sliding window is to estimate the middle value.
This delay in estimation is what gives coin-hopping miners the necessary time to mine many blocks while the difficulty is still low.
On the other hand, NEFDA shows negative correlation between consecutive hour-buckets which indicates that it responds rapidly to hash rate fluctuations.
More importantly, no positive feedback is present which is what is expected given the properties of history agnosticism and lack of autocorrelation.

\section{Conclusion}
\label{ref:conclusion}
We have showed how the behavior of economically rational miners can lead to severe instabilities in transaction throughput as a consequence of a positive feedback loop in {\tt cw-144}.
To mitigate periods of undesired (either too low or too high) transaction throughput, we derived NEFDA, a DA which does not lead to the formation of a positive feedback loop and can cope effectively with sudden hash rate fluctuations.
We explained how NEFDA exhibits desirable properties in the form of history agnosticism and lack of significant positive autocorrelation and demonstrated through simulations how NEFDA reduces target volatility, and in turn high variations in block solve times.
% Additionally, we show how NEFDA is configurable in the level of responsiveness by setting the smoothing factor parameter, i.e. higher for lower reactiveness and vice versa.
Ultimately, NEFDA constitutes a viable alternative for both large and small blockchains (in terms of baseline hash rate) when configured appropriately and may thus guarantee more stable transaction throughput.

% \section*{Acknowledgments}
% The authors would like to thank zawy12 for helpful discussions on difficulty algorithms based on exponential moving averages.
% The authors thank Daniel Perez for helpful technical support.
% Sam Werner and Dragos Ilie are supported by funding from the Brevan Howard Centre for Financial Analysis.

\bibliographystyle{abbrv}
\bibliography{references}

\end{document}